\documentclass[copyright,creativecommons,noderivs,noncommercial]{eptcs}
\providecommand{\event}{}
\providecommand{\volume}{2}
\providecommand{\anno}{2009}
\providecommand{\firstpage}{83}
\usepackage{latexsym}
\usepackage[english]{babel}
\usepackage[dvips]{graphicx}
\usepackage{color}
\usepackage{enumerate}
\usepackage{multirow}
\usepackage{psfrag}
\usepackage{pstricks, pst-node, pst-tree}
\usepackage{subfigure, graphicx, epsfig}
\usepackage{xspace}
\usepackage{amsmath, amsthm, amssymb}
\usepackage{url}
\usepackage{listings}
\usepackage{fancyvrb}
\usepackage{breakurl}

\def\qed{\hfill{\qedboxempty}      
  \ifdim\lastskip<\medskipamount \removelastskip\penalty55\medskip\fi}

\def\qedboxempty{\vbox{\hrule\hbox{\vrule\kern3pt
                 \vbox{\kern3pt\kern3pt}\kern3pt\vrule}\hrule}}

\def\qedfull{\hfill{\qedboxfull}   
  \ifdim\lastskip<\medskipamount \removelastskip\penalty55\medskip\fi}

\def\qedboxfull{\vrule height 4pt width 4pt depth 0pt}

\newtheorem{exampleAux}{Example}[]

\definecolor{gray}{rgb}{0.85,0.85,0.85}

\newcommand {\indigolog} {\textsf{IndiGolog}\xspace}
\newcommand {\congolog} {\textsf{ConGolog}\xspace}

\newcommand {\IndiGolog} {\textsf{IndiGolog}\xspace}

\newcommand {\mobidis} {\textsf{SmartPM}\xspace}

\newcommand{\manets}{\textsc{manet}s\xspace}

\newcommand {\myi}{\emph{(i)}~}
\newcommand {\myii}{\emph{(ii)}~}
\newcommand {\myiii}{\emph{(iii)}~}
\newcommand {\myiv}{\emph{(iv)}~}

\title{Adaptive Process Management in Highly Dynamic and Pervasive
Scenarios}
\author{Massimiliano de Leoni
\institute{Dipartimento di Informatica e Sistemistica\\
SAPIENZA -- Universit\`a  di Roma}
\email{deleoni@dis.uniroma1.it}}

\def\titlerunning{Adaptive PM in Highly Dynamic and Pervasive Scenarios}
\def\authorrunning{M.~de Leoni}

\begin{document}
\maketitle

\begin{abstract}
Process Management Systems (PMSs) are currently more and more used
as a supporting tool for cooperative processes in pervasive and
highly dynamic situations, such as emergency situations, pervasive
healthcare or domotics/home automation. But in all such situations,
designed processes can be easily invalidated since the execution
environment may change continuously due to frequent unforeseeable
events. This paper aims at illustrating the theoretical framework
and the concrete implementation of \mobidis, a PMS that features a
set of sound and complete techniques to automatically cope with
unplanned exceptions. PMS \mobidis is based on a general framework
which adopts the Situation Calculus and \indigolog.
\end{abstract}

\section{Introduction}

Nowadays organisations are always trying to improve the performance
of the processes they are part of. It does not matter whether such
organisations are dealing with classical static business domains,
such as loans, bank accounts or insurances, or with pervasive and
highly dynamic scenarios. The demands are always the same: seeking
more efficiency for their processes to reduce the time and the cost
for their execution.

According to the definition given by the Workflow Management
Coalition,\footnote{\url{http://wfmc.org}} a workflow is ``the
computerised facilitation of automation of a business process, in
whole or part''. The Workflow Management Coalition defines a
Workflow Management System as ``a system that completely defines,
manages and executes workflows through the execution of software
whose order of execution is driven by a computer representation of
the workflow logic''. Workflow Management Systems (WfMSs) are also
known as Process Management Systems (PMSs), and we are going to use
both of them interchangeably throughout this thesis. Accordingly,
this thesis uses many times word ``process'' is place of word
``workflow'', although the original acceptation of the former is not
intrinsically referring to its computerised automation.

%

In this paper 
we turn our
attention to highly dynamic and pervasive scenarios. Pervasive
scenarios comprise, for instance, emergency management, health care
or home automation (a.k.a. domotics). All of these scenarios are
characterised as being very dynamic and turbulent and subject to an
higher frequency of unexpected contingencies with respect to
classical scenarios. Therefore, PMSs for pervasive scenarios should
provide a higher degree of operational flexibility/adaptability.

According to Andresen and Gronau~\cite{A_G@InfoRes} adaptability can
be seen as an ability to change something to fit to occurring
changes. Adaptability is to be understood here as the ability of a
PMS to adapt/modify processes efficiently and fast to change
circumstances. Adaptation aims at reducing the gap of the
\emph{virtual reality}, the (idealized) model of reality that is
used by the PMS to deliberate, from the \emph{physical reality}, the
real world with the actual values of conditions and
outcomes~\cite{GiacomoRS98}. Exogenous events may make deviate the
virtual reality from the physical reality. The reduction of this gap
requires sufficient knowledge of both kinds of realities (virtual
and physical). Such knowledge, harvested by the services performing
the process tasks, would allow the PMS to sense deviations and to
deal with their mitigation.

In pervasive settings, efficiency and effectiveness when carrying on
processes are a strong requirement. For instance, in emergency
management saving minutes could result in saving injured people,
preventing buildings from collapses, and so on. Or, pervasive
health-care processes can cause people's permanent diseases when not
executed by given deadlines. In order to improve effectiveness of
process execution, adaptation ought to be as automatic as possible
and to require minimum manual human intervention. Indeed, human
intervention would cause delays, which might not be acceptable.

In theory there are three possibilities to deal with deviations:
\begin{enumerate}
    \item Ignoring deviations -- this is, of course, not feasible in general,
    since the new situation might be such that the PMS is no more able to
    carry out the process instance.
    \item Anticipating all possible discrepancies  -- the idea is to
    include in the process schema the actions to cope with each of
    such failures. This can be seen as a
    \texttt{try-catch} approach, used in some programming languages such as
    Java.
    The process is defined as if exogenous actions cannot occur, that
    is everything runs fine (the \texttt{try} block). Then, for each
    possible exogenous event, a \texttt{catch} block is designed in
    which the method is given to handle the corresponding exogenous
    event.
    For simple and mainly static
    processes, this is feasible and valuable; but, especially in mobile
    and highly dynamic scenarios, it is quite impossible to take
    into account all exception cases.
    \item Devising a general recovery method able to handle any
    kind of exogenous events -- considering again the metaphor of
    try/catch, there exists just one \texttt{catch} block, able to
    handle any exogenous events, included the unexpected.
    The \texttt{catch} block activates the general recovery method to modify
    the old process $P$ in a process $P'$ so that $P'$
    can terminate in the new environment
    and its goals are included in those of $P$.
    This approach relies on the execution monitor (i.e., the module intended for execution
    monitoring) that detects discrepancies leading the process instance not
    to be terminable. When they are sensed,
     the control flow moves to the \texttt{catch} block.
    An important challenge here is to build the monitor which
    is able to identify which exogenous events are relevant, i.e.~that make impossible process to terminate, as well as to \emph{automatically}
    synthesize $P'$ during the execution
    itself.
\end{enumerate}

\begin{table}
  \centering{
  \caption{Adaptability in the leading PMSs (as from~\cite{deLeoniPhD}).}
  \begin{small}
\begin{tabular}{|p{0.15\columnwidth}|c|c|c|}
\hline
\textbf{Product} & \textbf{Manual} & \textbf{Pre-planned} & \textbf{Unplanned} \\
\hline YAWL &  & \checkmark & \\
\hline COSA & \checkmark & \checkmark & \\
\hline Tibco & \checkmark & \checkmark & \\
\hline WebSphere & \checkmark & \checkmark & \\
\hline SAP & \checkmark & \checkmark  & \\
\hline OPERA & \checkmark & \checkmark & \\
\hline ADEPT2  & \checkmark &  & \\
\hline ADOME & \checkmark &  & \\
\hline AgentWork & \checkmark &  & \\
\hline
\end{tabular}
  \end{small}}
  \label{tab:adaptPMS}
  \end{table}

Table~\ref{tab:adaptPMS} shows the adaptability features of the most
valuable PMSs according to the state-of-art analysis described
in~\cite{deLeoniPhD}. Column \textbf{Manual} refers to the
possibility of a responsible person who manually changes the process
schema to deal with exogenous events. Column \textbf{Pre-planned}
concerns the feature of defining policies to specify the adaptation
behaviour to manage some exogenous events, whose possible occurrence
is foreseeable a priori. The last column \textbf{Unplanned} refers
to the third approach in the classification above.

The third approach seems to be the most appropriate when dealing
with scenarios where \myi the frequency of unexpected exogenous
events are relatively high and \myii there are several exogenous
events that cannot be foreseen before their actual occurrence.
Unfortunately, as the table shows, the world leading PMSs are unable
to feature the third approach.

This paper describes \mobidis, a PMS that features some sound and
complete techniques according to the third approach described above.
Such techniques are meant to improve the degree of \emph{automatic}
adaptation to react to very frequent changes in the execution
environment and fit processes accordingly. The techniques proposed
here are based on Situation Calculus~\cite{ReiterBook} and automatic
planning, conceived to coordinate robots and intelligent agents. The
concrete implementation, namely \mobidis, is based on the
\indigolog\ interpreter developed at University of Toronto and RMIT
University, Melbourne.

In \mobidis, every entity performing task is generally named
``service''. A service may be a human actor/process participant as
well as an automatic service that execute a certain job (e.g., a
SOAP-based Web Service).

Let us consider a scenario for emergency management where processes
show typical a complexity that is comparable to business settings.
Therefore, the usage of PMS is valuable to coordinate the activities
of emergency operators. In these scenarios, operators are typically
equipped with low-profile devices, such as PDAs, which several
services are installed on. Such services may range from usual
GUI-based applications to automatic ones. For instances, some
applications can be installed to fill questionnaires or take
pictures. In addition, PDAs can be provided with some automatic
services that connect to the Civil Protection headquarters to
retrieve information for the assessment of the affected area and
possibly send back the data collected.

PDAs communicate with each other by Mobile Ad-hoc Networks
(\manets), which are Wi-Fi networks that do not rely on a fixed
infrastructure, such as Access Points. Devices can be the final
recipients of some packets sent by other devices as well as they can
act as relays and forward packets towards the final destination.

In order to orchestrate the services installed on operator devices,
such devices need to be continually connected to the PMS through a
loose connection: devices and the PMS can communicate if there
exists a path of nodes that connects them in the graph of the
communication links.

In the virtual reality, devices are supposed to be continuously
connected (i.e., a path always exists between pairs of nodes). But
in this physical reality continuous connections cannot be
guaranteed: the environment is highly dynamic and the movement of
nodes (that is, devices and related operators) within the affected
area, while carrying out assigned tasks, can cause disconnections
and make deviate the two reality. Disconnections results in the
unavailability of nodes and, hence, the services provided. From the
collection of actual user requirements~\cite{H_C_dL_M_M_B_S@HCI09},
it results that typical teams are formed by a few nodes (less than
10 units), and therefore frequently a simple task reassignment is
not feasible. Indeed, there may not be two ``similar'' services
available to perform a given task. Reordering task executions would
not solve the problem, either. There is no guarantee that eventually
those services that provide unique capability connect again to the
PMS.

So, adaptaption is needed: adaptability might consist in this case
to recover the disconnection of a node X, and that can be achieved
by assigning a task ``Follow X'' to another node Y in order to
maintain the connection. When the connection has been restored, the
process can progress again.

\section{Preliminaries}
\label{sec:basics}

In this section we introduce the Situation Calculus, which we use to
formalize \mobidis and its adaptation features. The Situation
Calculus \cite{ReiterBook} is a second-order logic targeted
specifically for representing a dynamically changing domain of
interest (the world). All changes in the world are obtained as
result of \emph{actions}. A possible history of the actions is
represented by a \emph{situation}, which is a first-order term
denoting the current situation of the world. The constant $s_0$
denotes the initial situation. A special binary function symbol
$do(\alpha,s)$ denotes the next situation after performing the
action $\alpha$ in the situation $s$. Action may be parameterized.

Properties that hold in a situation are called \emph{fluents}. These
are predicates taking a situation term as their last argument. For
instance, we could define the fluent $free(x,s)$ stating whether the
object $x$ is free in situation $s$, meaning no object is located on
$x$ in situation $s$.

Changes in fluents (resulting from executing actions) are specified
through \emph{successor state axioms}. In particular for each fluent
$F$ we have a successor state axioms as follows:
\[F(\overrightarrow{x},do(\alpha,s)) \Leftrightarrow
\Phi_F( \overrightarrow{x},do(\alpha,s),s)\]
where $\Phi_F(\overrightarrow{x},do(\alpha,s),s)$ is a formula with
free variables $\overrightarrow{x}$, $\alpha$ is an action, and $s$
is a situation. %

In order to control the executions of actions we make use of high
level programs expressed in \indigolog~\cite{S_DG_L_L@AMAI04}, which
is equipped with primitives for expressing concurrency.
Table~\ref{tab:IndiGolog} summarizes the constructs of \indigolog
used in this work. Basically, these constructs allow to define every
well-structured process as defined in \cite{KiepuszewskiHB00}. The
last table column shows the corresponding statement defined in the
\indigolog\ platform developed at University of Toronto and RMIT
University.\footnote{Downloadable at
\url{http://www.cs.toronto.edu/cogrobo/main/systems/index.html}}

\begin{table}[t]
\centering{ 
  \caption{\indigolog constructs.}\label{tab:IndiGolog}
\begin{small}
\begin{tabular}{|l|p{8cm}|p{4.3cm}|}
  \hline
  \textbf{Construct} & \textbf{Meaning} & \textbf{Platform Statement}\\\hline
  $a$ & A primitive action & \texttt{a}\\\hline
  $\phi?$ & Wait while the $\phi$ condition is false & \texttt{?(phi)} \\\hline
  $(\delta_1;\delta_2)$ & Sequence of two sub-programs $\delta_1$
  and $\delta_2$ & \texttt{[delta1,delta2]} \\\hline
  $proc~P(\overrightarrow{v})~\delta$ & Invocation of a procedure
  passing a vector $\overrightarrow{v}$ of parameters & \texttt{proc(P,delta)}\\\hline
  $(\phi;\delta_1) | (\neg\phi;\delta_2)$ & Exclusive choice between $\delta_1$ and $\delta_2$
  according to the condition $\phi$ & \texttt{ndet([?(phi);delta1],} \texttt{[?(neg(phi)),delta2])} \\\hline
  $while~\phi~do~\delta$ & Iterative invocation of $\delta$ & \texttt{while(phi,delta)}\\\hline
  $(\delta_1\parallel\delta_2)$ & Concurrent execution & \texttt{rrobin(delta1,delta2)}\\\hline
  $\delta^*$ & Indeterministic iteration of program execution (The platform statement
  limits the maximum iterations number to \texttt{n}) & \texttt{star(delta,n)}
  \\\hline
  $\Sigma (\delta)$ & Emulating off-line execution & \texttt{searchn(delta,n)}\\\hline
  $\pi a.\delta$ & Indeterministic choice of argument $a$ followed by the execution of $\delta$ & \texttt{pi(a,delta)} \\\hline
\end{tabular}
\end{small}
}
\end{table}

From the formal point of view, \indigolog programs are terms. The
execution of \congolog programs is expressed through a
\emph{transition semantic} based on single steps of execution. At
each step a program executes an action and evolves to a new program
which represents what remains to be executed of the original
program. Formally two predicates are introduced to specify such a
sematic:
\begin{itemize}
    \item $Trans(\delta',s',\delta'',s'')$, given a program $\delta'$
    and a situation $s'$, returns \myi a new situation $s''$ resulting from executing
    a single step of
    $\delta'$, and \myii $\delta''$ which is the remaining program to be executed.
    \item $Final(\delta',s')$ returns true when the program $\delta'$
    can be considered successfully completed in situation $s'$.
\end{itemize}

By using $Trans$ and $Final$ we can define a predicate
$Do(\delta',s',s'')$ that represent successful complete executions
of a program $\delta'$ in a situation $s'$, where $s''$ is the
situation at the end of the execution of $\delta'$. Formally:
\begin{displaymath}
Do(\delta',s',s'')\Leftrightarrow\exists\delta''.Trans^*(\delta',s',\delta'',s'')
\wedge Final(\delta'',s'')
\end{displaymath}
\noindent where $Trans^*$ is the definition of the reflective and
transitive closure of \emph{Trans}.

To cope with the impossibility of backtracking actions executed in
the real world, \indigolog\ incorporates a new programming
construct, namely the {\em search operator}. Let $\delta$ be any
\indigolog\ program, which provides different alternative executable
actions. When the interpreter encounters program $\Sigma(\delta)$,
before choosing among alternative executable actions of $\delta$ and
possible picks of variable values, it performs reasoning in order to
decide for a step which still allows
the rest of $\delta$ to terminate successfully. 
If $\delta$ is the entire program under consideration,
$\Sigma(\delta)$ emulates complete off-line execution.

\section{General Framework}\label{sec:GeneralFramework}

The general framework which we shall introduce in this paper is
based on the \textit{execution monitoring} scheme as described in
\cite{GiacomoRS98} for situation calculus agents. As we will later
describe in more details, when using \indigolog\ for process
management, we take tasks to be predefined sequences of actions (see
later) and processes to be \indigolog\ programs.
After each action, the PMS may need to align the internal world
representation (i.e., the virtual reality) with the external one
(i.e., the physical reality).

\begin{figure*}[]
\centering{
 \includegraphics[width=0.6\columnwidth]{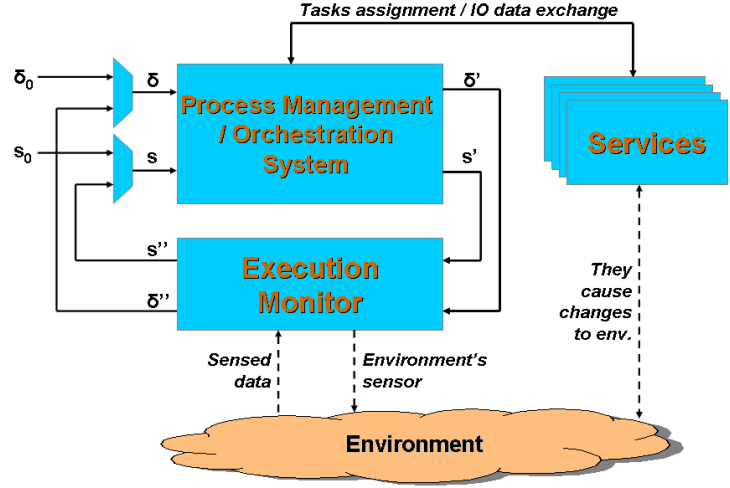}
} \caption{Execution Monitoring.} \label{fig:monitoring}
\end{figure*}

Before a process starts, PMS takes the initial context from the real
environment and builds the corresponding initial situation $S_0$, by
means of first-order logic formulas. It also builds the program
$\delta_{0}$ corresponding to the process to be carried on.
Then, at each execution step, PMS, which has a complete knowledge of
the internal world (i.e., its virtual reality), assigns a task to a
service. The only ``assignable'' tasks are those whose preconditions
are fulfilled. A service can collect data required needed to execute
the task assigned from PMS. When a service finishes executing a
task, it alerts PMS of that.

The execution of the PMS can be interrupted by the \textit{monitor}
module when a misalignment between the virtual and the physical
realities is discovered. In that case, the monitor \textit{adapts}
the (current) program to deal with such discrepancy.

In Figure~\ref{fig:monitoring}, the overall framework is depicted.
At each step, the PMS advances the process $\delta$ in situation $s$
by executing an action, resulting then in a new situation $s'$ with
the process $\delta'$ remaining to be executed. Both $s'$ and
$\delta'$ are given as input to the monitor, which also collects
data from the environment through \emph{sensors}.\footnote{Here, we
refer as \emph{sensors} not only proper sensors (e.g., the ones
deployed in sensor networks), but also any software or hardware
component enabling to retrieve contextual information. For instance,
it may range from GIS clients to specific hardware that makes
available the communication distance of a device to its
neighbors.~\cite{dL_M_R@WETICE07}}
If a discrepancy between the virtual reality as represented by $s'$
and the physical reality is sensed, then the monitor changes $s'$ to
$s''$, by generating a sequence of actions that explains the changes
perceived in the environment, thus re-aligning the virtual and
physical realities.
Notice, however, that the process $\delta'$ may \textit{fail} to
execute successfully (i.e., assign all tasks as required) in the new
(unexpected) situation $s''$. If so, the monitor adapts also the
(current) process by performing suitable recovery changes and
generating then a new process $\delta''$. At this point, the PMS is
resumed and the execution continues with program-process $\delta''$
in situation $s''$.

\section{Process Formalisation in Situation Calculus}
\label{sec:BPMFormalization}

Next we detail the general framework proposed above by using
Situation Calculus and \indigolog. We use some domain-independent
predicates to denote the various objects of interest in the
framework:
\begin{itemize}
    \item $service(a)$: $a$ is a service
    \item $task(x)$: $x$ is a task
    \item $capability(b)$: $b$ is a capability
    \item $provide(a,b)$: the service $a$ provides the capability $b$
    \item $require(x,b)$: the task $x$ requires the capability $b$
\end{itemize}
In the light of these predicates, we have defined a shortcut to
refer to the capability of a certain service $a$ to perform a list
of tasks, a.k.a.~worklist. Service $a$ can execute a certain
worklist $wrkList$ iif $a$ provides all capabilities required by all
tasks in the worklist:
\[
Capable(a,wrklist) \Leftrightarrow \big( \forall b,t. t \in wrkList
\wedge require(b,t) \Rightarrow provide(a,b) \big)
\]

Every task execution is the sequence of four PMS actions: \myi the
assignment of the task to a service, resulting in the service being
not free anymore; \myii the notification to the service to start
executing the task. Then, the service carries out the tasks and,
after receiving the service notification of the task conclusion,
\myiii the PMS acknowledges the successful task termination.
Finally, \myiv the PMS releases the service, which becomes free
again. We formalise these four actions as follows:
\begin{itemize}
    \item $Assign(a,x)$: task $x$ is assigned to a service $a$
    \item $Start(a,x,p)$: service $a$ is allowed to
    start the execution of task $x$. The input provided is $p$.
    \item $AckTaskCompletion(a,x)$: service $a$ concluded successfully the
    executing of $x$.
    \item $Release(a,x)$: the service $a$ is released with respect
    to task $x$.
\end{itemize}
In addition, services can execute two actions:
\begin{itemize}
    \item $readyToStart(a,x)$: service $a$ declares to be ready to
    start performing task $x$
    \item $finishedTask(a,x,q)$: service $a$ declares to have completed
    executing task $x$ returning output $q$.
\end{itemize}

The terms $p$ and $q$ denote arbitrary sets of input/output, which
depend on the specific task. Special constant $\emptyset$ denotes
empty input or output.

The interleaving of actions performed by the PMS and services is as
follows. After the assignment of a certain task $x$ by
$Assign(a,x)$, when the service $a$ is ready to start executing, it
executes action $readyToStartTask(a,x)$. At this stage, PMS executes
action $Start(a,x,p)$, after which $a$ starts executing task $x$.
When $a$ completes task $x$, it executes the action
$finishedTask(a,x,q)$. Specifically, we envision that actions
$finishedTask(\cdot)$ are those in charge of changing properties of
world as result of executing tasks. When $x$ is completed, PMS is
allowed in any moment to execute sequentially
$AckTaskCompletion(a,x)$ and $Release(a,x)$. The program coding the
process will the executed by only one actor, specifically the PMS.
Therefore, actions $readyToStartTask(\cdot)$ and
$finishedTask(\cdot)$ are considered as external and, hence, not
coded in the program itself.

For each specific domain, we have several fluents representing the
properties of situations. Some of them are modelled independently of
the domain whereas others, the majority, are defined according to
the domain. If they are independent of the domain, they can be
always formulated as defined in this chapter. Among the
domain-independent ones, we have fluent $free(a,s)$, that denotes
the fact that the service $a$ is free, i.e., no task has been
assigned to it, in the situation $s$. The corresponding successor
state axiom is as follows:
\begin{equation}\label{eq:freeAxiom}
\begin{array}{l}
free(a,do(t,s)) \Leftrightarrow {}\\
\qquad\big(\forall x.t \neq Assign(a,x) \wedge free(a,s) \big) \vee {}\\
\qquad\big( \neg free(a,s) \wedge \exists x.t = Release(a,x) \big)
\end{array}
\end{equation}
This says that a service $a$ is considered free in the current
situation if and only if $a$ was free in the previous situation and
no tasks have been just assigned to it, or $a$ was not free and it
has been just released. There exists also the domain-independent
fluent $enabled(x,a,s)$ which aims at representing whether service
$a$ has notified to be ready to execute a certain task $x$ so as to
enabled it. The corresponding successor-state axiom:
\begin{equation}\label{eq:enabledAxiom}
\begin{array}{l}
enabled(x,a,do(t,s)) \Leftrightarrow \\\qquad\big( enabled(x,a,s)
\wedge \forall q. t \neq finishedTask(a,x,q) \big)\vee
\\ \qquad\big(  \neg enabled(x,a,s) \wedge t=readyToStartTask(a,x) \big)
\end{array}
\end{equation}
This says that $enabled(x,a,s)$ holds in the current situation if
and only if it held in the previous one and no action
$finishedTask(a,x,q)$ has been performed or it was false in the
previous situation and $readyToStartTask(a,x)$ has been executed.
This fluent aims at enforcing the constraints that the PMS can
execute $Start(a,x,p)$ only after $a$ performed $begun(a,x)$ and it
can execute $AckTaskCompletion(a,x,q)$ only after
$finishedTask(a,x,q)$. This can represented by two pre-conditions on
actions $Start(\cdot)$ and $AckTaskCompletion(\cdot)$:
\begin{equation}\label{eq:possStartStop}
\begin{array}{l}
\forall p.Poss(Start(a,x,p),s) \Leftrightarrow enabled(x,a,s) \\
\forall p.Poss(AckTaskCompletion(x,a),s) \Leftrightarrow \neg
enabled(x,a,s)
\end{array}
\end{equation}
provided that $AckTaskCompletion(x,a)$ never comes before
$Start(x,a,p),s$.

Furthermore, we introduce a domain-independent fluent
$started(x,a,p,s)$ that holds if and only if an action
$Start(a,x,p)$ has been executed but the dual
$AckTaskCompletion(x,a)$ has not yet:
\begin{equation}\label{eq:startedAxiom}
\begin{array}{l}
started(a,x,p,do(t,s)) \Leftrightarrow \\\qquad\big(
started(a,x,p,s) \wedge t \neq Stop(a,x) \big)\vee
\\
\qquad\big( \nexists p'.started(x,a,p',s) \wedge t=Start(a,x,p)
\big)
\end{array}
\end{equation}

In addition, we make use, in every specific domain, of a predicate
$available(a,s)$ which denotes whether a service $a$ is available in
situation $s$ for tasks assignment. However, $available$ is
domain-dependent and, hence, requires to be defined specifically for
every domain. Knowing whether a service is available is very
important for the PMS when it has to perform assignments. Indeed, a
task $x$ is assigned to the best service $a$ which is available and
provides every capability required by $x$. The fact that a certain
service $a$ is free does not imply it can be assigned to tasks
(e.g., in the example described above it has to be free as well as
it has to be indirectly connected to the coordinator). The
definition of $available(\cdot)$ must enforce the following
condition:
\begin{equation}\label{eq:availableAxiom}
\forall a~s. available(a,s) \Rightarrow free(a,s)
\end{equation}

We do not give explicitly pre-conditions to task. We assume tasks
can always be executed. We assume that, given a task, if some
conditions do not hold, then the outcomes of that tasks are not as
expected (in other terms, it fails).

\section{The \mobidis System}\label{sec:architecture}

\begin{figure*}[t!]
\centering{
 \includegraphics[width=0.6\columnwidth]{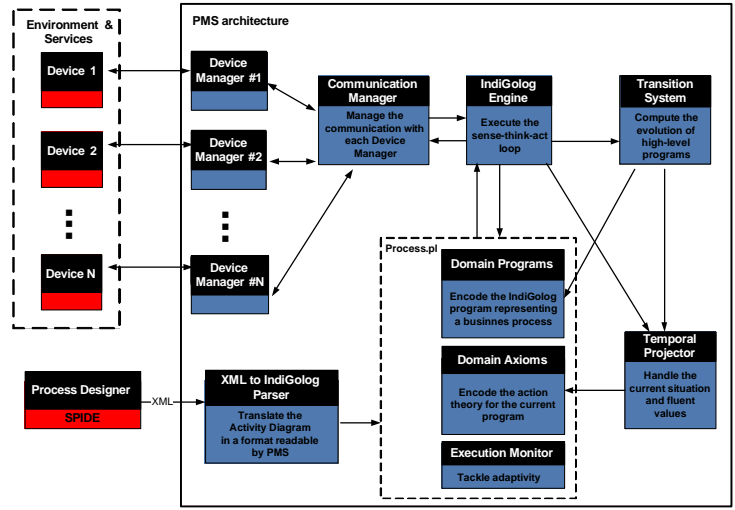}
} \caption{Architecture of the PMS.} \label{fig:fig_architecture}
\end{figure*}

This section aims at describing the internal structure of PMS.
Figure~\ref{fig:fig_architecture} shows its conceptual architecture.
At the beginning, a responsible person designs an Activity Diagram
through SPIDE, a \emph{Process Designer} Graphical tool with
which \mobidis is equipped. Later, Such a tool translates the
Activity Diagram in a XML format file. Then, such a XML file is
loaded into PMS. The \emph{XML-to-\indigolog\ Parser} component
translates this specification in a \emph{Domain Program}, the
\indigolog\ program corresponding to the designed process, and a set
of \emph{Domain Axioms}, which is the action theory that comprises the
initial situation, the set of available actions with their pre- and
post-conditions.

When the program is translated in the Domain Program and Axioms, a
component named \emph{Communication Manager} (CM) starts up all of
\emph{device managers}, which are basically some drivers for making
communicate PMS with the services and sensors installed on devices.
For each real world device PMS holds a device manager. Each device
manager is also intended for notifying the associated device about
every action performed by the \mobidis engine as well as for
notifying the \mobidis engine about the actions executed by the
services of the associated device.

After this initialization process, CM activates the
\emph{\indigolog\ Engine}, which is in charge of executing
\indigolog\ programs. Then, CM enters into a passive mode where it
is listening for messages arriving from the devices through the
device managers. In general, a message can be a exogenous event
harvested by a certain sensor installed on a given device as well as
a message notifying the start or completion of a certain task. When
CM judges a message as significant, it forwards it to \indigolog.
For instance, relevant messages may be signals of the task
completion or the sudden unavailability of a given device.

In sum, CM is responsible of deciding which device should
perform certain actions, instructing the appropriate device
managers to communicate with the device services and collecting the
corresponding sensing outcome. The \indigolog\ Engine is intended to
execute a \textit{sense-think-act} interleaved
loop~\cite{Kowalski95}. The cycle repeats at all times the following
three steps:
\begin{enumerate}
\item check for exogenous events that have occurred;
\item calculate the next program step; and
\item if the step involves an action, \textit{execute} the action,
instructing the Communication Manager.
\end{enumerate}

The \indigolog\ Engine relies on two further modules named
\emph{Transition System} and \emph{Temporal Projector}. The former
is used to compute the evolution of \indigolog\ programs according
to the statements' semantic, whereas the latter is in charge of
holding the current situations throughout the execution as well as
letting evaluate the fluent values for taking the right decision of
the actions to perform.

The last module that is worth mentioning is the \emph{Execution
Monitor} (MON), which get notifications of exogenous events from the
Communication Manager. It decides whether adaptation is needed and
adapts accordingly the process. Section~\ref{sec:mobidisMonitoring}
gives some additional details of the concrete implementation of
monitoring and adaptation.

\section{A Concrete Example from Emergency Management}

\begin{figure}[t!] \centering
  \includegraphics[height=0.6\textheight]{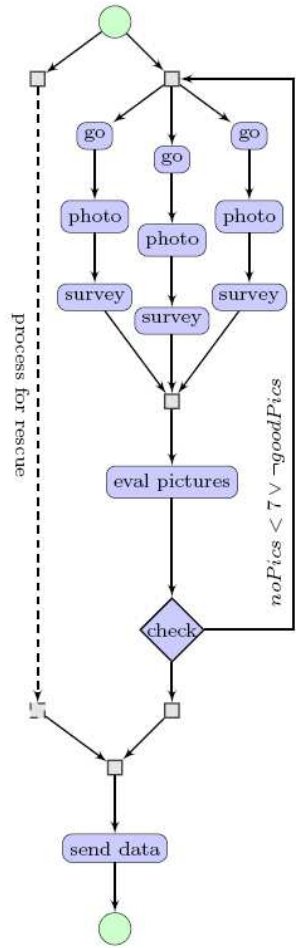}
\caption{An activity diagram of a process concerning emergency
management.} \label{fig:pms}
\end{figure}

\begin{figure}[t!] \centering
\begin{minipage}{0.5\textwidth}
\begin{scriptsize}
\begin{verbatim}
proc(main,
 prioritized_interrupts(
  [interrupt(exogEvent, monitor),
   interrupt(true, process),
   interrupt(neg(finished), wait)]
)).
proc(process, [rrobin(processRescue,
  while(or(noPhotos<7,neg(goodPics)),
   [rrobin(
        [manageTasks(
            [workitem((go,id19,loc(5,5)),
             workitem((photo,id20,loc(5,5)),
             workitem((survey,id21,loc(5,5))]),
         manageTasks(
            [workitem((go,id19,loc(15,15)),
             workitem((photo,id20,loc(15,15)),
             workitem((survey,id21,loc(15,15))]),
         manageTasks(
            [workitem((go,id19,loc(50,50)),
             workitem((photo,id20,loc(50,50)),
             workitem((survey,id21,loc(50,50))]),
        ]
    ),
    manageTasks([workitem((evalPics,id28,input)])
   ]) % end of while
 ), % end concurrent subprocesses
 manageTasks([workitem((sendData,id29,input)])
]).

proc(manageTasks(WrkList),
  pi(srvc,
    [?(and(Available(srvc),Capable(srvc,WrkList))),
      manageExecution(WrkList,srvc),
    ]
)).
proc(manageExecution([],Srvc),[]).
proc(manageExecution([workitem(Task,Id,I)|TAIL],Srvc),
     [assign(Task,Id,Srvc,I),
     start(Task,Id,Srvc,I),
     ackTaskCompletion(Task,Id,Srvc),
     release(Task,Id,Srvc,I),
     manageExecution(TAIL,Srvc)
     ]
    )
\end{verbatim}
\end{scriptsize}
\end{minipage}
\caption{An example of process management with \IndiGolog.}
\label{fig:pmsIndiGolog}
\end{figure}

We turn to describe the approach by an example concerning emergency
management in an area affected by an earthquake. The emergency
response process in question comprises various activities that may
need to be adapted on-the-fly to react to unexpected exogenous
events that could arise during the operation. Figure~\ref{fig:pms}
depicts an Activity Diagram of a process consisting of two
concurrent branches; the final task is \emph{send data} and can only
be executed after the branches have successfully completed. The left
branch, abstracted out from the diagram, is built from several
concurrent processes involving tasks \emph{rescue},
\emph{evacuation} and others. The right branch begins with the
concurrent execution of three sequences of tasks: \emph{go},
\emph{photo}, and \emph{survey}. When all survey tasks have been
completed, the task \emph{evaluate pictures} is executed. Then, a
condition is evaluated on the resulting state at a decision point
(i.e., whether the pictures taken are of sufficient quality). If the
condition holds, the right branch is considered finished; otherwise,
the whole branch should be repeated.

Figure \ref{fig:pmsIndiGolog} shows some parts of the \IndiGolog\
program representing the process of the example. The code proposes
here has been slightly simplified and abstracted for the sake of
brevity. The main procedure, called \texttt{main}, involves three
interrupts running at different priorities. The first highest
priority interrupt fires when an exogenous event occurs (i.e.,
condition \texttt{exogEvent} is true). In such a case, the
\texttt{monitor} procedure is executed, evaluating whether or not
adaptation is required (see Section~\ref{sec:mobidisMonitoring}).

If no exogenous event has occurred, the second interrupt triggers
and execution of the actual emergency response process is attempted.
Procedure \texttt{process}, also shown in the figure, encodes the
Activity Diagram of the example process. It relies, in turn, on
procedure \texttt{manageTasks(WrkLists)}, where \texttt{WrkLists} is
a sequence of elements \texttt{workitem(T,I,D)}, each one
representing a task \texttt{T}, with identifier \texttt{I}, and
input data \texttt{D}, which needs to be performed. This procedure
is meant to manage the execution of all tasks in the worklist, and
it assigns them all to a \emph{single} service that provides every
capability required.

Of course, to assign tasks to an service, \mobidis needs to reason
about the available ones, their current state (e.g., their
location), and their capabilities, as not every service is capable
of performing any task. In fact, before assigning the first task in
any task list, procedure \texttt{manageTasks(WrkLists)} executes a
\emph{pick} operation is done to choose a Service \texttt{srvc} that
is involved in no task execution (i.e., fluent \texttt{Free(actr)}
holds) and able to execute the whole worklist.

Once a suitable service has been chosen, PMS assigns the list of
tasks to it by executing \linebreak \texttt{assign(srvc,WrkList)}.
In addition to inform the service about the task assignment, such an
action turns fluent \texttt{Free(actr)} to false.

Then, PMS calls procedure \texttt{manageExecution(WrkList)}, which
handles the execution of each task in the list. For each task T in
the list (with identifier \texttt{I} and input data \texttt{D}), the
procedure invokes action \texttt{start(T,D,I,srvc)} that provides
the required information to the chosen service \texttt{srvc}. In
this way, the service is instructed to begin working on the task and
receives the required input. When a service finishes executing an
assigned task, it alerts \mobidis via action
\texttt{finishedTask(T,srvc)}; PMS acknowledges by performing
\texttt{ackTaskCompletion (T,D,actr)}. When the whole work-item list
is execution, the PMS releases the service by executing the action
\texttt{release(T,D,actr)}, after which fluent \texttt{Free(srvc)}
is turned to true again.

It is worth mentioning that, if the process being carried out cannot
execute temporarily further, the lowest priority interrupt fires.
This interrupt makes PMS wait for the conditions in which some tasks
can be executed. The fact that the process gets stuck does not imply
necessarily the occurrence of some relevant exogenous events. It
could be also caused by the fact that next tasks can be only
assigned to services that are currently busy busy performing other
tasks. The latter situation does not prevent processes from being
completed successfully; indeed, such services will be eventually
free to work on those tasks.

\section{Adaptation in \mobidis}

\subsection{Monitoring Formalisation} \label{sec:BPMAdaptiveness}

Next we formalize how the monitor works. Intuitively, the monitor
takes the current program $\delta'$ and the current situation $s'$
from the PMS's virtual reality and, analyzing the physical reality
by sensors, introduces fake actions in order to get a new situation
$s''$ which aligns the virtual reality of the PMS with sensed
information. Then, it analyzes whether $\delta'$ can still be
executed in $s''$, and if not, it adapts $\delta'$ by generating a
new correctly executable program $\delta''$. Specifically, the
monitor work can be abstractly defined as follows (we do not model
how the situation $s''$ is generated from the sensed information):

\begin{equation}
\begin{array}{l}
Monitor(\delta',s',s'',\delta'') \Leftrightarrow 
\big(Relevant(\delta', s', s'') \wedge Recovery(\delta', s', s'',
\delta'') \big) \vee {}\\ \qquad \big(\neg Relevant( \delta', s',
s'') \wedge \delta'' = \delta' \big)
\end{array}
\label{equ:monitor}
\end{equation}

\noindent where: \myi $Relevant(\delta',s',s'')$ states whether the
change from the situation $s'$ into $s''$ is such that $\delta'$
cannot be correctly executed anymore; and \myii $Recovery(\delta',
s', s'', \delta'')$ is intended to hold whenever the program
$\delta'$, to be originally executed in situation $s'$, is
adapted to $\delta''$ in order to be executed in situation
$s''$.

Formally $Relevant$ is defined as follows:

\begin{displaymath}
Relevant(\delta',s',s'') \Leftrightarrow \neg
SameConfig(\delta',s',\delta',s'')
\end{displaymath}

\noindent where $SameConfig(\delta',s',\delta'',s'')$ is true if
executing $\delta'$ in $s'$ is ``equivalent'' to executing
$\delta''$ in $s''$ (see later for further details).

In this general framework we do not give a definition for
$SameConfig(\delta',s',\delta'',s'')$. However we consider any
definition for $SameConfig$ to be correct if it denotes a
bisimulation \cite{MilnerBook}. Formally, for every $\delta',
s',\delta'',s''$ holds:

\begin{enumerate}
    \item $Final(\delta',s') \Leftrightarrow Final(\delta'',s')$
    \item
    $\forall~a,\delta'.Trans\big(\delta',s',\overline{\delta'},do(a,s')\big) \Rightarrow$
    \\$\exists~\overline{\delta''}.Trans\big(\delta'',s'',\overline{\delta'},do(a,s'')\big)
    \wedge SameConfig\big(\overline{\delta'},do(a,s),\overline{\delta''},do(a,s'')\big)$
    \item
    $\forall~a,\delta'.Trans\big(\delta'',s'',\overline{\delta'},do(a,s'')\big) \Rightarrow$
    \\$\exists~\overline{\delta''}.Trans\big(\delta',s',\overline{\delta'},do(a,s')\big)
    \wedge SameConfig\big(\overline{\delta''},do(a,s''),\overline{\delta'},do(a,s')\big)$
\end{enumerate}

Intuitively, a predicate $SameConfig(\delta',s',\delta'',s'')$ is
said to be correct if $\delta'$ and $\delta''$ are terminable either
both or none of them. Furthermore, for each action $a$ performable
by $\delta'$ in the situation $s'$, $\delta''$ in the situation
$s''$ has to enable the performance of the same actions (and
viceversa). Moreover, the resulting configurations
$(\overline{\delta'},do(a,s'))$ and $(\overline{\delta''},do(a,s'))$
must still satisfy $SameConfig$.

The use of the bisimulation criteria to state when a predicate
$SameConfig(\cdots)$ is correct, derives from the notion of
equivalence introduced in \cite{HiddersDAHV05}. When comparing the
execution of two formally different business processes, the internal
states of the processes may be ignored, because what really matters
is the process behavior that can be observed. This view reflects the
way a PMS works: indeed what is of interest is the set of tasks that
the PMS offers to its environment, in response to the inputs that
the environment provides.

Next we turn our attention to the procedure to adapt the process
formalized by $Recovery(\delta,s,s',\delta')$. Formally is defined
as follows:

%

\begin{equation}
\begin{array}{l}
Recovery(\delta',s',s'',\delta'') \Leftrightarrow {}
\exists \delta_a,\delta_b.\delta''=\delta_a;\delta_b \wedge
Deterministic(\delta_a) \wedge \\\quad Do(\delta_a,s'',s_b) \wedge
SameConfig(\delta',s',\delta_b,s_b)
\end{array}
\label{equ:recovery}
\end{equation}

$Recovery$ determines a process $\delta''$ consisting of a
\emph{deterministic} $\delta_a$ (i.e., a program not using the
concurrency construct), and an arbitrary program $\delta_b$. The aim
of $\delta_a$ is to lead from the situation $s''$ in which
adaptation is needed to a new situation $s_b$ where
$SameConfig(\delta',s',\delta_b,s_b)$ is true.

The nice feature of \textsc{Recovery} is that it asks to search for
a linear program that achieves a certain formula, namely
$SameState(s',s'')$. That is we have reduced the synthesis of a
recovery program to a classical Planning problem in AI
\cite{TraversoBook}. As a result we can adopt a well-developed
literature about planning for our aim. In particular, if the
services and input and output parameters are finite, then the
recovery can be reduced to \emph{propositional} planning, which is
known to be decidable in general (for which very well performing
software tools exists).

Notice that during the actual recovery phase $\delta_a$ we disallow
for concurrency because  we need full control on the execution of
each service in order to get to a recovered state. Then the actual
recovered program $\delta_b$ can again allow for concurrency.

In the previous sections we have provided a general description on
how adaptation can be defined and performed. Here we choose a
specific technique that is actually feasible in practice. Our main
step is to adopt a specific definition for $SameConfig$, here
denoted as \textsc{SameConfig}, namely:
\begin{equation}
\textsc{SameConfig}(\delta',s',\delta'',s'') \Leftrightarrow
SameState(s',s'') \wedge \delta'=\delta''
\label{equ:SameConfigConcrete}
\end{equation}

\indent In other words, \textsc{SameConfig} states that $\delta'$,
$s'$ and $\delta''$, $s''$ are the same configuration if \myi all
fluents have the same truth values in both $s'$ and $s''$
($SameState$), and \myii $\delta''$ is actually
$\delta'$.\footnote{Observe that $SameState$ can actually be defined
as a first-order formula over the fluents, as the conjunction of
$F(s') \Leftrightarrow F(s'')$ for each fluent $F$.} In
papers~\cite{deLeoniPhD,DBLP:conf/bpm/LeoniMG07}, we have proved
that the above-defined \textsc{SameConfig} is a correct
bisimulation.

Using Equation~\ref{equ:SameConfigConcrete} as $SameConfig$
definition feasible in practice, relevancy results to be:
\begin{equation}
\begin{array}{l}
\textsc{Relevant}(\delta',s',s'') \Leftrightarrow \neg
SameState(s',s'')
\end{array}
\label{equ:RelevantConcrete}
\end{equation}
In the next section, we are going to show how the abstract planner
specification given here has been concretely used inside \mobidis.
Specifically the current version of \mobidis uses the proportional
planner available in the \indigolog platform developed by University
of Toronto and RMIT in Melbourne. In order to adapt, \mobidis is
based on the concrete definitions of relevancy and $SameConfig$
given by Equations~\ref{equ:RelevantConcrete}
and~\ref{equ:SameConfigConcrete}.

\subsection{The Execution Monitoring and Adaptation}
\label{sec:mobidisMonitoring}

\begin{figure}[t!] \centering
\begin{minipage}{0.7\textwidth}
\begin{Verbatim}[fontsize=\scriptsize]
proc(monitor,[ndet(
            [?(neg(relevant))],
            [?(relevant),recovery]
          )]).

proc(recovery,  searchn([searchProgram],10).

proc(searchProgram, [star(pi([Task,Id,Input,srvc],
    [?(and(Available(srvc),
            Capable(srvc,[workitem(Task,Id,Input)]))),
      manageExecution([workitem(Task,Id,Input)],srvc)])),
    ?(SameState)]).
\end{Verbatim}
\end{minipage}
\caption{The procedure for managing automatic adaptation with the
\indigolog interpreter.} \label{fig:adaptIndiGolog}
\end{figure}

As already told, adaptation amounts to find a linear program (i.e.,
without concurrency) that is meant to be ``appended'' before the
current \indigolog program remaining to be executed. Such a linear
program is meant to resolve the gap that was just sensed by
restoring the values of affected fluents to those before the
occurrence of the deviation.

Figure~\ref{fig:adaptIndiGolog} shows how adaptability has been
concretely implemented in \mobidis. The execution of the process
being carried out by \mobidis can be interrupted by the
\texttt{monitor} procedure when a misalignment between the virtual
and the physical reality is discovered.

The \texttt{monitor} procedure is the concrete coding of
Equation~\ref{equ:monitor} and relies on procedure
\texttt{relevant}. Procedure \texttt{relevant} returns true if the
exogenous event has created a gap between the physical and virtual
reality that is in accord with Equation~\ref{equ:RelevantConcrete}.
For this aim, \mobidis keeps a ``copy'' of the expected value of
each defined fluent so that when an exogenous action is sensed it
can check whether the action has altered the value of some fluent.

If the gap is relevant, procedure \texttt{recovery} is invoked. It
amounts to find a linear program (i.e., without concurrency) to
reduce the gap sensed as well as, if such a program is found, to
execute it. After executing such a linear program, the program coded
by routine \texttt{process} (and its possible sub-routines) can
progress again. This behaviour is equivalent to that expressed
formally in Equation~\ref{equ:recovery} where the adapting linear
program is ``appended before'' and, hence, executed before the
remaining process.

The \texttt{recovery} procedure looks for a sequence of actions that
brings to a situation in which procedure \texttt{SameState} returns
true: $\Sigma \big((\pi a.a)^*; SameState?\big)$. Procedure
\texttt{SameState} tests whether executing $(\pi a.a)^*$ really has
really reduced the gap. The use of the \indigolog's lookahead
operator $\Sigma$ guarantees the action sequence $(\pi a.a)^*$ is
chosen so as to make \texttt{SameState} true. In fact, we do not
look for any action sequence $(\pi a.a)^*$ but we reduce the search
space since we search for sequences of invocations of procedure
\texttt{manageExecution} with appropriate parameters.

\section{Conclusion}

Most of existing PMSs are not completely appropriate for very
dynamic and pervasive scenarios. Indeed, such scenarios are
turbulent and subject to a higher frequency of unexpected
contingencies with respect to usual business settings that show a
static static and foreseeable behaviour. This paper describes
\mobidis, an adaptive PMS that is able to adapt processes thus
recovering from exceptions. Adaptation is synthesized automatically
without relying either on the intervention of domain experts or on
the existence of specific handlers planned in advance to cope with
specific exceptions. Space limitation has prevented from including
concrete examples of adaptation: interested readers can refer
to~\cite{deLeoniPhD}.

Future works aim mostly at integrating \mobidis with state-of-art
planners. Indeed, current implementation relies on the \indigolog
planner, which performs a blind search without using smarter
techniques recently proposed to reduce the search space by removing
a priori all the possibility surely taking to no solution. The most
challenging issue is to convert Action Theories and \indigolog\
programs in a way they can be given as input to planners (e.g.,
converting to PDDL~\cite{F_L@JAIR07}).

\section*{Acknowledgments} The author wishes to thank to Giuseppe De Giacomo, Andrea
Marrella, Massimo Mecella and Sebastian Sardina, who have
contributed to different aspects of the \mobidis development.
\bibliographystyle{eptcs}
\bibliography{LiteratureReview,aPMS,mypapers,Implementation}

\end{document}